**Room temperature single-photon detectors for high bit rate quantum key distribution**

L. C. Comandar[1,2], B. Fröhlich [1,a)], M. Lucamarini[1], K. A. Patel[1,2], A. W. Sharpe[1], J. F. Dynes[1], Z. L. Yuan[1], R. V. Penty[2] and A. J. Shields[1]

[1]Toshiba Research Europe Ltd, 208 Cambridge Science Park, Milton Road, Cambridge CB4 0GZ, United Kingdom

[2]Engineering Department, Cambridge University, 9 J J Thomson Ave, Cambridge CB3 0FA, United Kingdom

We report room temperature operation of telecom wavelength single-photon detectors for high bit rate quantum key distribution (QKD). Room temperature operation is achieved using InGaAs avalanche photodiodes integrated with electronics based on the self-differencing technique that increases avalanche discrimination sensitivity. Despite using room temperature detectors, we demonstrate QKD with record secure bit rates over a range of fiber lengths (e.g. 1.26 Mbit/s over 50 km). Furthermore, our results indicate that operating the detectors at room temperature increases the secure bit rate for short distances.

---

[a)] Author to whom correspondence should be addressed. Electronic mail: bernd.frohlich@crl.toshiba.co.uk

Quantum key distribution (QKD) allows the distribution of secret digital keys on optical fibers, the security of which is related to the laws of physics.[1-5] Measurement of the quantum states used to carry the key bits by a third party will cause errors to their encoding, allowing legitimate users to detect any eavesdropping attempt. QKD has been demonstrated to achieve Mbit/s secure key rates[6,7] and long communication distance.[8] Recent research has focused on extending the practical applicability of QKD through, e.g., multiplexing with conventional data signals,[9,10] multi-user access networks,[11] or improving the security level and efficiency by developing advanced protocols.[12,13]

However, one practical limitation present in many demonstrations is that cooling is required for the detectors, either cryogenic for those based on superconducting nanowires[8] and superconducting transition edge-sensors,[14] or thermo-electrical for InGaAs avalanche photodiodes (APDs).[6,7] A photon detector operating at room temperature is highly desirable, as it will reduce the system complexity and footprint. Up-conversion detectors use the superior single-photon counting properties of Si APDs and can operate at room temperature,[15,16] however, the up-conversion step adds a significant layer of complexity. Certain types of QKD systems such as continuous-variable QKD do not rely on single-photon counting and use uncooled detectors. However, this benefit is largely outweighed by the need for more demanding post-processing[17] leading to limitations on the achievable transmission distance and secure key rates that are orders of magnitude lower than for QKD based on single-photon detectors.

Significant advances have been made for single-photon detection technologies based on APDs. Techniques such as self-differencing (SD),[18] sine-wave gating,[19-21] or harmonic subtraction[22] have enabled much improved single-photon discrimination sensitivity, lower dark count rates, faster timing resolution, higher detection efficiencies and higher maximum detection rates. With SD detectors, a detection rate of 1 GHz has been recently reported.[23] However, it still remains unknown whether the benefits of these new technologies will permit QKD over long distances and with high rates when the APD is operated at room temperature. Although the suitability of a room temperature sine gating detector for QKD is briefly discussed in Refs. [24,25], the communication distance is limited to 25 km and the achieved bit rate is orders of magnitude lower than in state-of-the-art QKD experiments with cooled detectors.[6]

Here, we report room temperature operation of InGaAs APD detectors enabled by an integrated SD circuit, and we implement them in a state-of-the-art QKD system. We show that these room temperature detectors do not degrade the system performance for a fiber distance up to 65 km. In fact, we achieve a record secure key rate of 1.26 Mbit/s for a fiber distance of 50 km and we still demonstrate a useable key rate at a fiber length of 100 km. Moreover, our investigation indicates that room temperature operation can even lead to superior performance of the detector compared to cooled operation for short fiber distances.

We start our investigation by measuring the detector system performance over a range of temperatures. The characterization setup is shown in Figure 1. A pulsed diode laser is synchronized to 1/64 of the APD gating frequency and has its output attenuated to an average photon number per pulse $\mu = 0.1$ before being coupled into the fiber pigtail of the diode. The diode is switched periodically above and below the breakdown voltage by a square wave signal with a peak-to-peak voltage of approximately 10 V. The temperature of the APDs is thermoelectrically controlled by a Peltier cooler in the range −30 to 20 °C. From the measured histograms at the output of the SD circuit we extract the two main error count sources, the dark count probability per gate $P_d$ and the afterpulse probability $P_a$, and the detector efficiency $\eta_{det}$ using a standard method.[18,26] The afterpulse probability is defined as the ratio of the afterpulse counts to the detected photon counts.

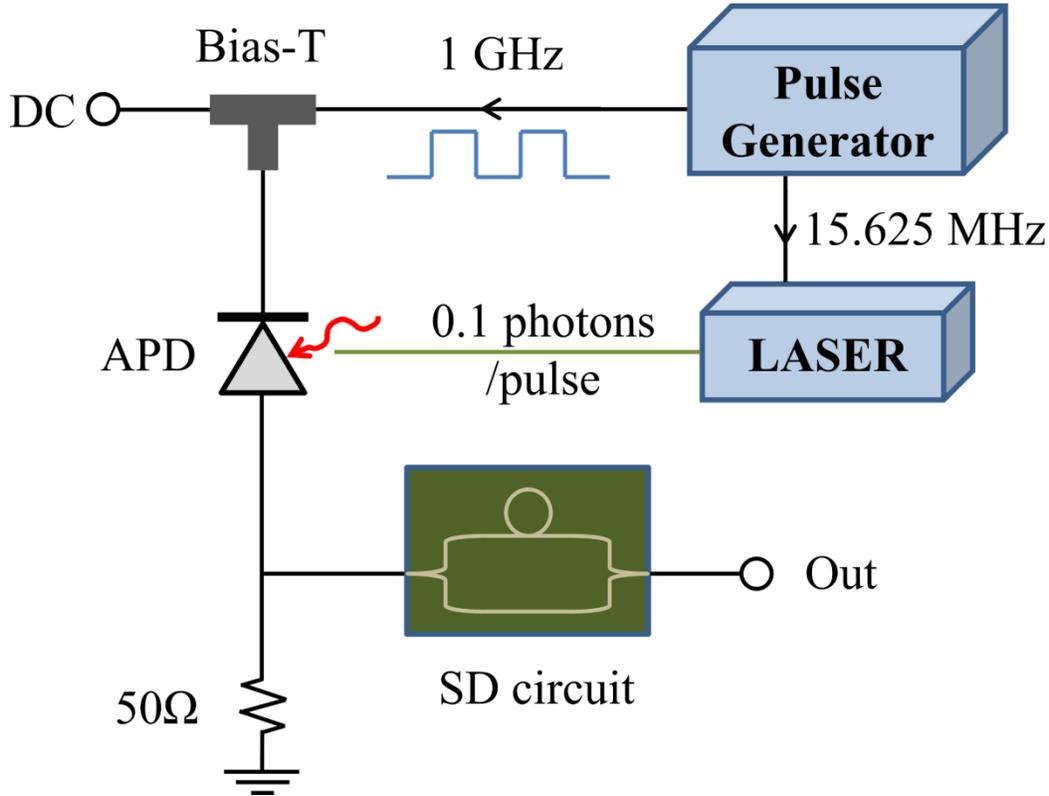

Fig. 1. Detector characterization setup including self-differencing (SD) circuit. A laser generates short light pulses with on average 0.1 photons/pulse which are synchronized with the gating signal of the detector. The avalanche signal of the gated APD is processed with the integrated SD circuit to enhance avalanche detector sensitivity.

In Figure 2 we compare the performance of a typical APD operating at −30 °C and at 20 °C (room temperature). Increasing the operating temperature has the effect of increasing the dark count rate and reducing the afterpulse probability.[27,28] For a detection efficiency of about 25% we find the dark count probability increases from $P_d = 3.1 \times 10^{-6}$ at −30 °C to $P_d = 5.9 \times$ $10^{-5}$ at 20 °C. Meanwhile, the afterpulse probability decreases from about 3.9% to 2.8% over the same temperature range. The timing jitter of approximately 60 ps does not vary noticeably between −30 °C and 20 °C. As shown below, the increase in dark count rate at room temperature reduces the maximum fiber length possible for QKD. However, it should be noted that for all but the longest fiber lengths, the afterpulse rate has a much larger effect upon the secure bit rate. Thus for short fiber lengths the secure bit rate increases when the detector is operated at room temperature.

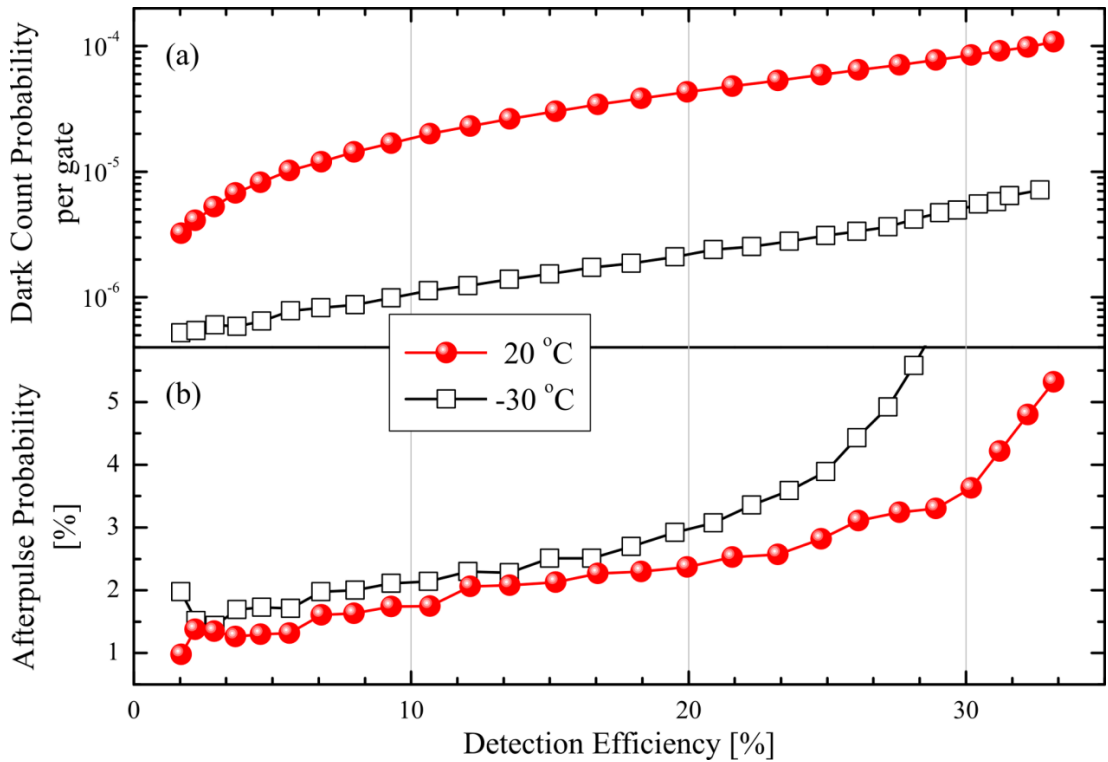

Fig. 2. Detector characterization results for a typical APD used for our measurements. (a) Measured dark count probabilities per gate and (b) afterpulse probabilities for the APD operating at −30 °C and at room temperature (20 °C).

We implement room temperature detectors into a state-of-the-art high bit rate QKD system.[6] The QKD system uses phase encoding and strongly attenuated laser pulses as the signal carriers. To maximize the secure bit rate we implement the decoy protocol using three different pulse intensities. The signal (μ=0.42 photons per pulse, sent with a probability of 98.83%), decoy (μ=0.042 photons per pulse, sent with a probability of 0.78%) and vacuum (μ=0.0007 photons per pulse, sent with a probability of 0.39%). The system implements an efficient version of the BB84 protocol with unequal bases probabilities of $p_z = 15/16$ and $p_x = 1/16$. The finite sample size effect is taken into account with a security parameter of $\varepsilon = 10^{-10}$, which is a measure of the probability of information leakage. The secure key is distilled from 20 minutes key sessions if not reported otherwise. Using InGaAs SD-APD single-photon detectors, thermoelectrically cooled to −30 °C, this system has previously delivered a secure key rate of 1.093 Mbit/s at a fiber distance of 50 km.[6] Here, we replace only the detectors in the setup and leave all other parts unchanged.

For a first test we choose a fiber distance of 50 km with a link loss of around 10.1 dB. The QKD system performance is measured over a range of detector temperatures between −30 °C and 20 °C, with the detector efficiency fixed at 25%. As shown in Fig. 3(a), the key rate at −30 °C is 1.34 Mbit/s, the highest reported to date at this fiber length. As the temperature is increased, there is only a small, if not negligible, variation in the secure bit rate. At room temperature, we achieve a secure bit rate of 1.26 Mbit/s, a mere 6% reduction from that measured at −30 °C. Note that this secure rate is even higher than previously reported with cooled detectors.[6] Hence, room temperature detectors are capable of supporting QKD at this distance with an increase in its overall performance. The reason is that at short distances the noise is dominated by afterpulses and these are reduced by a higher temperature.

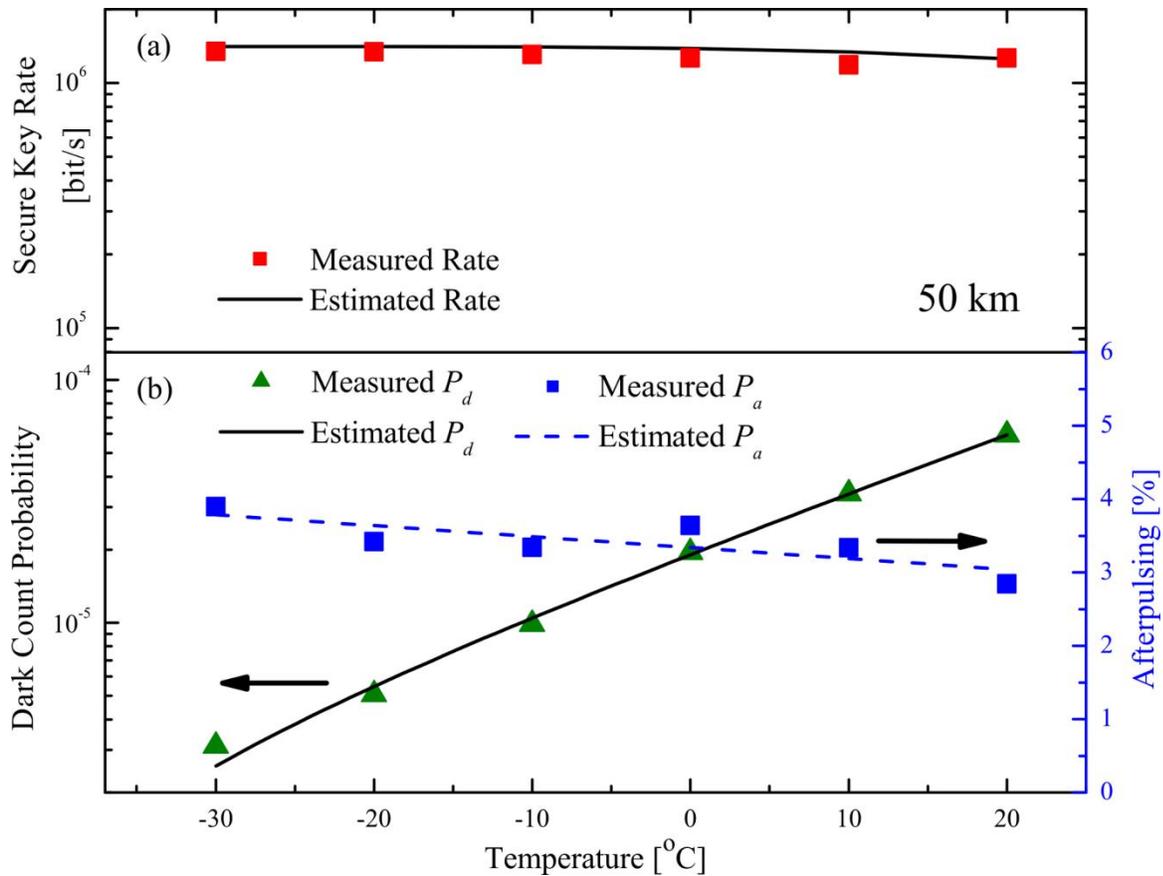

Fig. 3. (a) Measured secure key rate (solid squares) and simulated key rate (line) as a function of APD operating temperature at an efficiency of about 25% over 50 km fiber. (b) Measured and fitted dependency of dark count probability per gate ($P_d$) and afterpulsing probability ($P_a$) on temperature.

In Figure 3(b) this is shown more clearly. In fact, the dark count probability increases exponentially with temperature, while the afterpulsing probability decreases with temperature. We fit an exponential increase to the measured dark count probability and a linear decrease to afterpulsing from 3.89% at −30 °C to 2.82% at 20 °C. The fitted detector error count rates are then used to simulate the temperature dependence of the secure key rate. As shown in Fig. 3(a), the simulated secure bit rate (solid line) varies little over the entire temperature range and is in excellent agreement with the measured secure bit rate. The flat temperature response at this fiber distance is a result of the reduction in afterpulse noise with increasing temperature that largely cancels the effect of the increase in dark count noise.

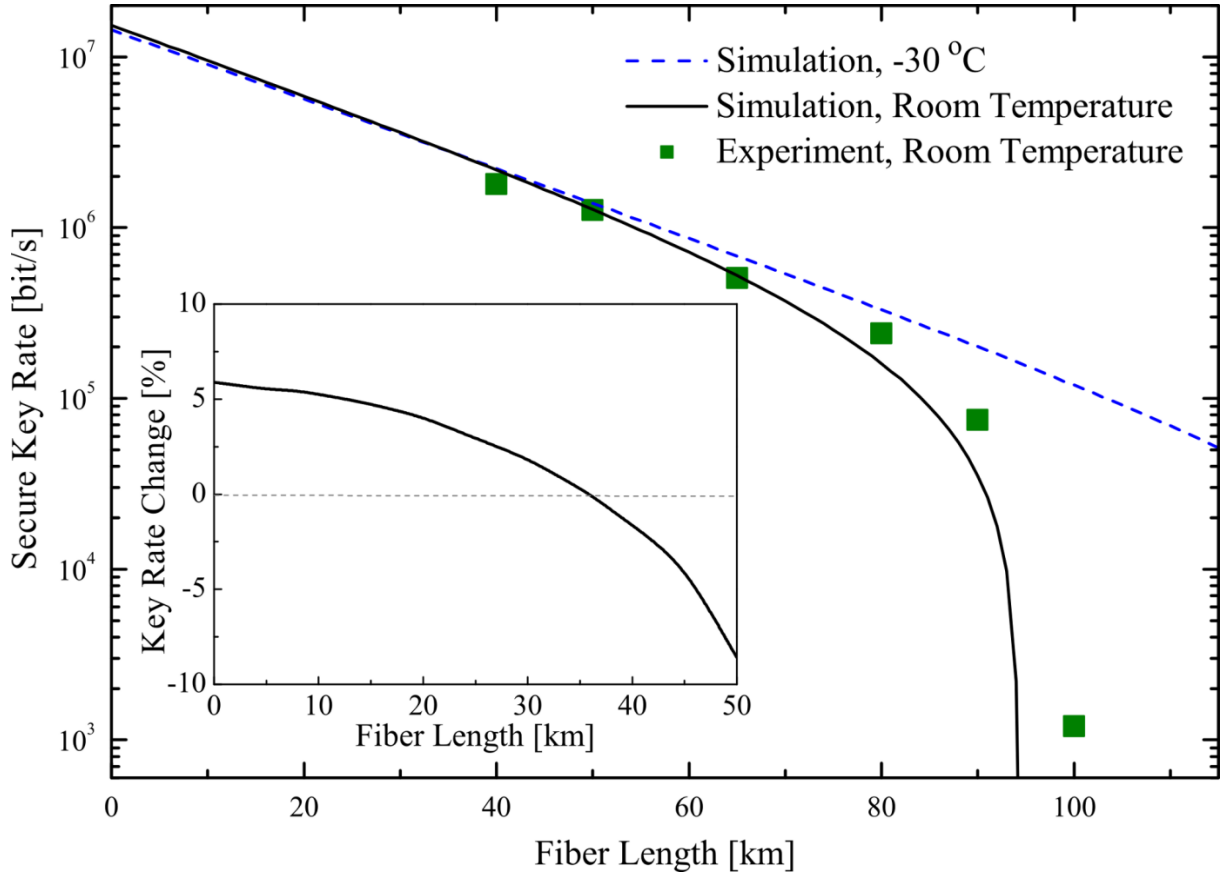

Fig. 4. Experimentally measured secure key rates (solid squares) with room temperature APDs as a function of fiber length. Also shown are theoretical simulations of the secure bit rates with the APDs operating at −30 °C (dashed line) and room temperatures (solid line) with detection efficiency fixed at 25%. Inset: relative change in the secure key rate between 20 °C and −30 °C detectors. The key rate change is defined as $(S_{20} - S_{-30})/S_{20}$ where S represents the secure key rate at a given temperature.

For shorter fiber lengths a higher secure bit rate can be obtained when using room temperature detectors compared to cooled operation. Figure 4 shows theoretical simulations and experimental data of the secure key rates for detectors operating at either room temperature or cooled to −30 °C, with detection efficiency set to 25%. For both temperatures the secure key rates decrease exponentially for most of the distance due to the fiber loss of 0.2 dB/km; then they fall sharply to zero at their respective cut-off lengths. This cut-off is a result of the dark counts exceeding the system error threshold. There is a cross-over in the theoretical curves at 35 km (Fig. 4, inset), below which higher secure bit rates are actually obtainable with room temperature detectors. At these short fiber distances, detector dark counts are negligibly low as compared to the photon detection rate even for room temperature detectors. Any reduction in the afterpulsing will therefore translate into an increase in the secure bit rate. For fiber lengths exceeding 50 km, the secure bit rate becomes noticeably lower for room temperature detectors, because the increased dark counts can be no longer compensated by the reduction in afterpulsing. The QKD distance is thus limited to about 95 km for room temperature detectors.

It is desirable to experimentally verify the QKD performance over various fiber distances. However, limited by the electronics, we can only demonstrate the full potential of the QKD performance for fiber distances of 40 km and longer. Shorter fiber distances will result in saturation of the photon count rate, and hence the secure bit rate. Experimental results for fiber lengths from 40 km and greater are shown in Figure 4. We obtain an average secure key rate of 1.79 Mbit/s, 1.26 Mbit/s and 507 kbit/s for fiber lengths of 40, 50, and 65 km, respectively. These bit rates represent the highest ever reported at their respective fiber lengths, even with the finite-size effect considered here.[6,13] For fiber lengths greater than 65 km, detector dark counts become significant. Hence, we choose to change the operating conditions of the detector to achieve a better trade-off between efficiency and dark counts which optimizes the secure key rate. This optimization has extended the QKD distance to 100 km. At 100 km, the QKD session time is increased to 60 min to compensate for the decreased count rate by fiber loss, which otherwise would lead to a greater reduction of the key rate due to the finite key size. Secure bit rates of 240, 74.8, and 1.2 kbit/s have been obtained for 80, 90, and 100 km, respectively.

The comparison of the expected secure bit rates in Figure 4 highlights that the reduction in secure bit rate using room temperature detectors is marginal for fiber lengths up to 65 km. That is because the single-photon detection rate is high for these fiber lengths and so the increased dark count noise under room temperature operation has little impact. Only for the longest fiber distances, where the single-photon detection rate approaches the dark count noise level, we measure a significant reduction in the secure bit rate. For these longer distances a change in ambient temperature affects the operation of the system noticeably if no temperature stabilization is implemented, whereas for short distances no control system is necessary. However, a control system stabilizing the temperature to close to 20° C still offers a significant reduction in size and power consumption compared to a system cooling below zero degrees.

In summary, we have presented the results of a temperature dependent characterization of an InGaAs APD with an integrated self-differencing circuit running at 1 GHz with reduced noise performance. Using detectors operating at room temperature (20 °C) we obtain secure key rates between 1.79 Mbit/s and 1.2 kbit/s for fiber lengths between 40 km and 100 km, respectively. We conclude that room temperature operation of the single-photon detectors is a viable solution for quantum key distribution. This will enable more compact and power efficient next-generation quantum key distribution systems. Our results are also important for other single-photon counting applications such as optical time-domain reflectometry[29] or laser ranging.[30]

**Acknowledgements**

L. C. Comandar and K. A. Patel acknowledge personal support via the EPSRC funded CDT in Photonics System Development.